\documentclass[
 reprint, amsmath,amssymb,aps,]{revtex4-1}

\usepackage{graphicx}
\usepackage{dcolumn}
\usepackage{bm}
\usepackage{epstopdf}
\usepackage{subcaption}
\usepackage{floatrow}
\usepackage{color}
\newfloatcommand{capbtabbox}{table}[][\FBwidth]
\usepackage[T1]{fontenc}
\usepackage[font=small,labelfont=bf,tableposition=top]{caption}

\DeclareCaptionLabelFormat{andtable}{#1~#2  \&  \tablename~\thetable}

\def\Journal#1#2#3#4{{#1} {\bf #2}, #3 (#4)}

\def\NIMA{{\em Nucl. Instrum. Methods} A}

\def\NDS{{\em Nucl. Data Sheets} } 

\def\PLB{{\em Phys. Lett.} B}
\def\PRP{{\em Phys. Rep.}}

\def\APJ{\em The Astrophys. J.}

\def\SCI{\em Science}

\def\PRD{{\em Phys. Rev.} D} 
\def\PRC{{\em Phys. Rev.} C}

\def\APJ{\em ApJ}

\newcommand{\iso}[2]{{\ensuremath{{}^{#2}\mathrm{#1}}}}
\newcommand{\nel}{\ensuremath{\nu_e}}

\begin{document}

\title{Calculated solar-neutrino capture rate for a radiochemical \iso{Tl}{205}-based
solar-neutrino detector}

\author{Joel Kostensalo}%
\email{joel.j.kostensalo@student.jyu.fi}  
\affiliation{Department of Physics, University of Jyvaskyla, Finland}

\author{Jouni Suhonen }%
 \email{jouni.t.suhonen@jyu.fi} 
\affiliation{Department of Physics, University of Jyvaskyla, Finland}

\author{K. Zuber}
\email{zuber@physik.tu-dresden.de}
 \affiliation{Institute for Nuclear and Particle Physics, TU Dresden,  01069 Dresden, Germany}




\date{\today}

\begin{abstract}
Radiochemical experiments for low-energy solar-neutrino detection have been making 
headlines by exploiting the isotopes \iso{Cl}{37} and \iso{Ga}{71}. Such a very 
low-threshold measurement of this type can also be performed using \iso{Tl}{205},
which has been considered for decades for this purpose. A unique feature of this
detector nucleus is the integration is the solar-neutrino flux over millions of years 
owing to its long-living daughter \iso{Pb}{205}. In this study we have
calculated for the first time the cross section for the charged-current
solar-neutrino scattering off \iso{Tl}{205}. Taking into account the 
solar-model-predicted neutrino fluxes and the electron-neutrino survival probabilities, 
a solar-neutrino capture rate of 62.2 $\pm 8.6$ SNU 
is determined, a value significantly smaller than in previous estimates.
\end{abstract}

\pacs{Valid PACS appear here}
\maketitle


\section{Introduction}
\label{sec:intro}
Neutrinos play a key role in several aspects of astroparticle and nuclear 
physics \cite{eji19}. From the astrophysical point of view solar neutrinos can be 
monitored in real-time measurements which allows to study neutrino properties and 
also stellar structure and evolution. To date real-time monitoring of various 
neutrino chain reactions has been done by the Super-Kamiokande, the Sudbury 
Neutrino Observatory (SNO), KamLAND, and especially Borexino. 
Borexino was able to perform a common global fit of all the observed reactions 
of pp, pep, $^7$Be and $^8$B in one detector \cite{ago18}.

An alternative method to the above-mentioned ones, used by the first solar-neutrino 
experiments, are the radio-chemical observations. These experiments employ the
charged-current neutrino-nucleus scattering reaction
\begin{equation}
\nel + (A,Z) \rightarrow e^- + (A,Z+1) 
\end{equation}
for solar-neutrino detection. This reaction has been used in the pioneering 
Homestake experiment using \iso{Cl}{37} as detector material \cite{cle98} and in this
experiment a deficit with respect to expectation was found.
First measurements of the fundamental pp-neutrinos were based on $^{71}$Ga 
(GALLEX, GNO, SAGE).
Several other nuclides, with different reaction thresholds, have been considered 
for more refined overall spectral analyses \cite{bah89}. A very interesting candidate 
of this type is \iso{Tl}{205}, which has a very low threshold for solar neutrinos.

\section{The $^{205}\textrm{Tl}$ reaction}\label{sec:reaction}

The dominant charged-current neutrino-nucleus reaction under discussion is 
\begin{equation}\label{eq:trans}
\iso{Tl}{205} (1/2^+) + \nel \rightarrow \iso{Pb}{205} (1/2^-)  + e^- \,,
\end{equation}
which feeds the first excited state of \iso{Pb}{205} at 2.33 keV and is of
first-forbidden non-unique type \cite{KostensaloPLB}. Only a tiny portion of the 
feeding goes to the $5/2^-$ ground state of \iso{Pb}{205}, the corresponding 
transition being first-forbidden unique \cite{KostensaloPLB,Suhonen2007}.
According to the current Atomic Mass Evaluation \cite{wan17} the $Q$-value is given 
by 50.6 $\pm$ 0.5 keV, which is so far the lowest threshold among radiochemical approaches 
for solar-neutrino detection. This results in a total threshold of about 53 keV for
the transition (\ref{eq:trans}). Furthermore, a unique feature of this reaction is
the possibility for long-term monitoring of the average solar-neutrino flux  
and hence the mean solar luminosity over the last 4.31 millions years due to the 
long half-life $1.73(7)\times 10^7$ yr of \iso{Pb}{205} \cite{kon04}. Hence, such 
a measurement could shed light on the long-term stability of the Sun and therefore 
on the stability of stars in general \cite{bah89}. 
 
First studies of the Tl experiment were done by \cite{fre76,fre88,kie88,hen88}  
which later became the LOREX experiment (LORandite EXperiment) \cite{pav88,pav12}. 
While several experimental aspects have already been addressed or have been worked on, 
the major remaining uncertainty is the cross section for this
reaction. Hence, it is essential to get a reliable estimate of this cross section
and this work reports on calculation of this important quantity using current 
state-of-the-art techniques. 



\section{Calculation of the $^{205}\textrm{Tl}$ cross section}\label{sec:system}

The calculations for the neutrino-nucleus cross section are based on
the Donnelly-Walecka method \cite{OConnell1972,Donnelly1979} for the treatment 
of semi-leptonic processes in nuclei. Details of the formalism as it is applied here 
can be found from \cite{Ydrefors12}. A streamlined version has also been given 
in the recent papers \cite{Kostensalo40ar,Kostensalo71Ga}.

\begin{figure}
\begin{center}
\includegraphics[width=\textwidth]{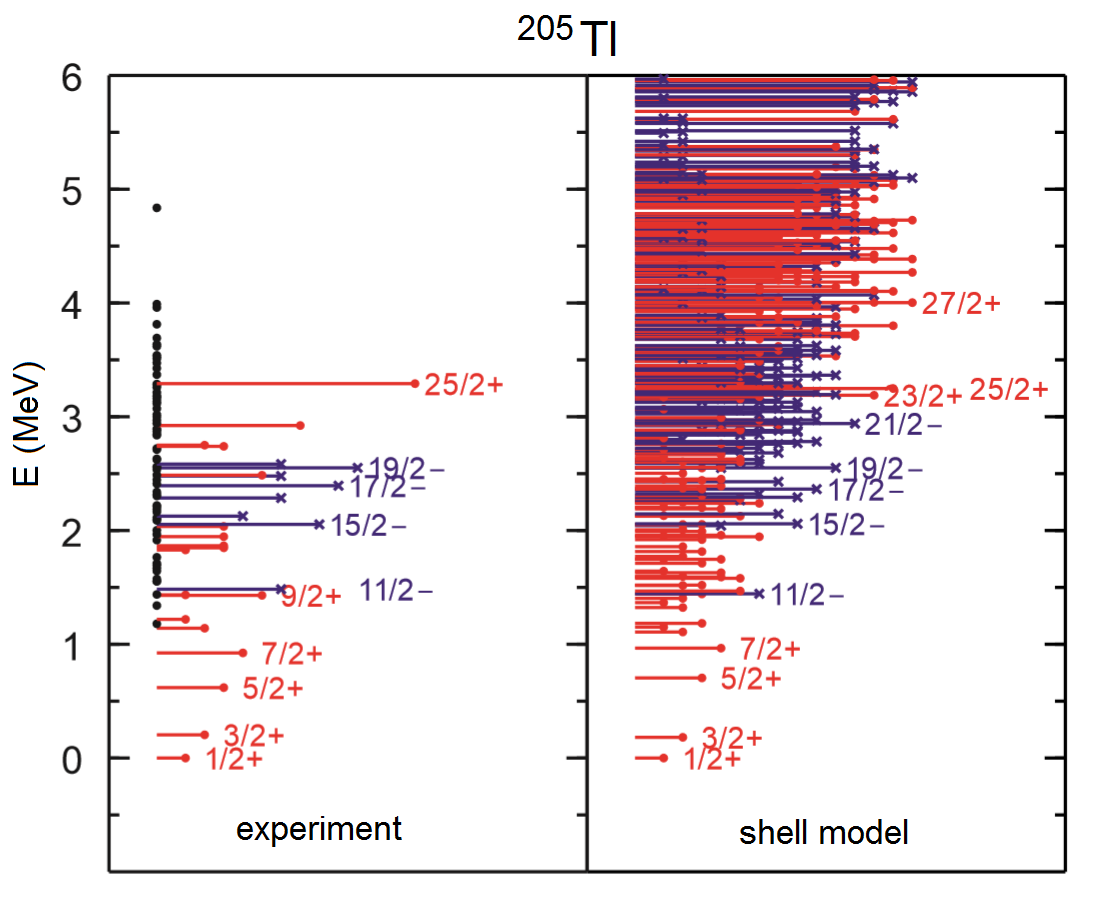}
\protect\caption[]{\footnotesize{Experimental and shell-model excitation spectra 
for \iso{Tl}{205}. Each horizontal bar represents a nuclear state and its length 
is proportional to the angular momentum of the state. \label{fig:tl}}}
\end{center}
\end{figure}

\begin{figure}
\begin{center}
\includegraphics[width=\textwidth]{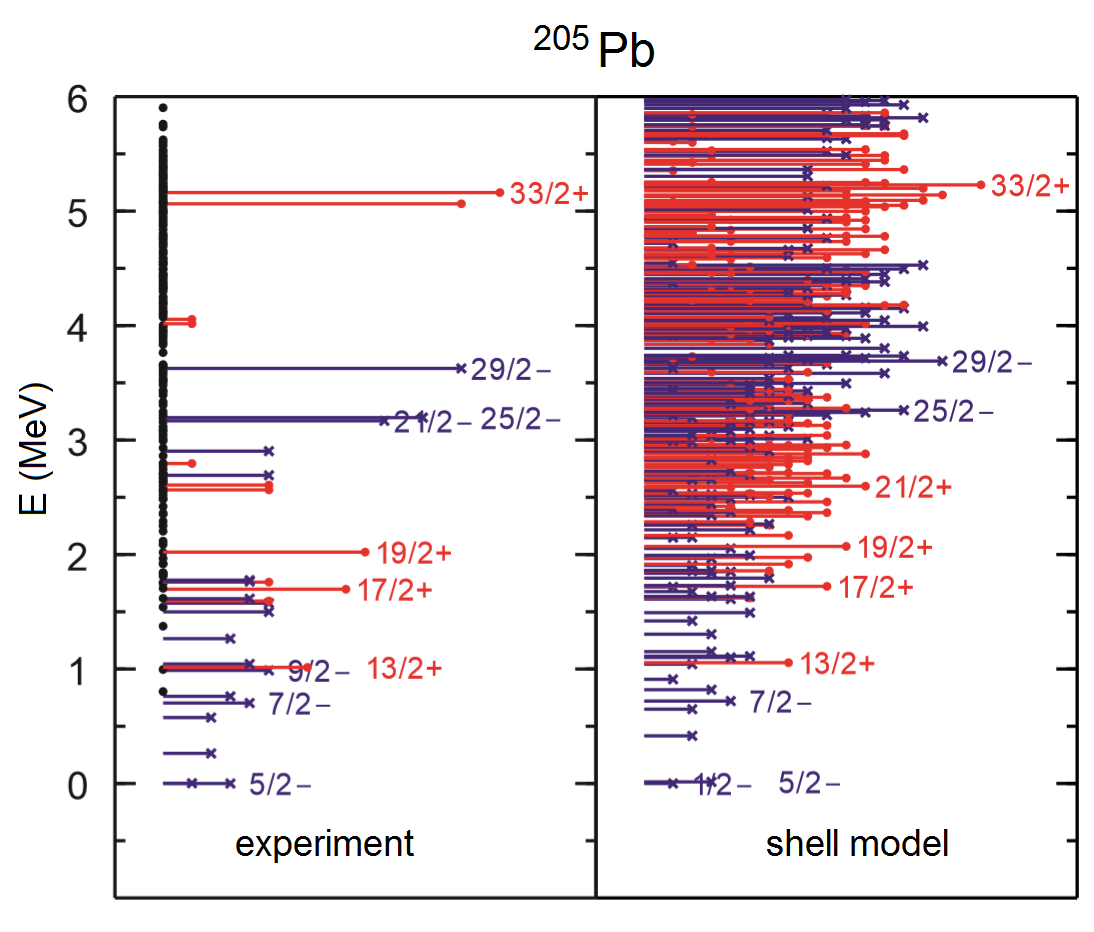}
\protect\caption[]{\footnotesize{Experimental and shell-model excitation spectra 
for \iso{Pb}{205}. Each horizontal bar represents a nuclear state and its length 
is proportional to the angular momentum of the state. \label{fig:pb}}}
\end{center}
\end{figure}

The nuclear-structure calculations were done in the shell-model framework using 
the shell-model code NuShellX@MSU \cite{nushellx} with the Hamiltonian 
khhe \cite{Warburton91} in the complete valence space spanned by the proton 
orbitals $0g_{7/2}$, $1d$, $2s$, and $0h_{11/2}$, and the neutron orbitals 
$0h_{9/2}$, $1f$, $2p$, and $0i_{13/2}$. As can be seen from Figures \ref{fig:tl} and
\ref{fig:pb} the energy spectra of the relevant nuclei are reproduced 
astonishingly well. The half-life of $^{205}$Pb is reproduced when $g_{\rm A}=0.75$ 
is adopted as the effective axial-vector coupling. This result is quite consistent 
with the previous calculations for beta decays in heavy nuclei \cite{Suhonen2017b}. 
In order to estimate the uncertainties related to nuclear structure, we consider 
here the range $g_{\rm A}=0.75$--$1.00$ for all transitions: a range which covers 
typical values adopted in large-scale shell-model calculations.

Since the exact energy of the low-lying states plays a significant role in 
determining the cross section for neutrinos with low energies, such as $pp$ 
neutrinos, the energies of the dominating low-lying states were adjusted to their 
experimental values. The energy-adjusted states were the lowest two $1/2^-$ states 
and the lowest three $3/2^-$ and $1/2^+$ states. Based on the ordering of the levels 
in the shell-model calculation the state at 803 keV was taken to be $1/2^-$ and 
the state at 996 keV to be $3/2^-$.

\begin{figure}
\begin{center}

\includegraphics[scale=0.50]{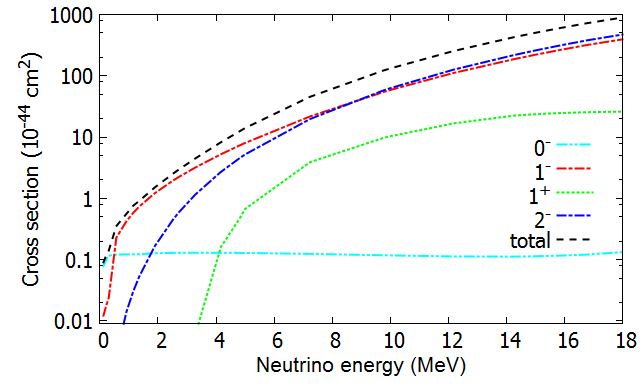}
\protect\caption[]{\footnotesize{Capture cross section for neutrino capture on 
\iso{Tl}{205} as function of neutrino energy for the leading multipolarities with 
$g_{\rm A}=1.00$. \label{fig:cmdiagram}}}
\end{center}
\end{figure}

\begin{figure}
\begin{center}
\includegraphics[width=\textwidth]{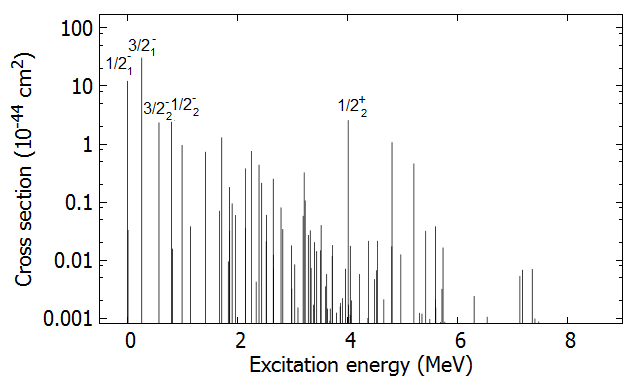}
\protect\caption[]{\footnotesize{Contributions of the individual states to the 
$^8$B neutrino cross section with $g_{\rm A}$=1.00. The horizontal axis gives
the excitation energy in \iso{Pb}{205}. \label{fig:contr}}}
\end{center}

\end{figure}

The contributions by multipolarity are shown in Fig.~\ref{fig:cmdiagram} and by 
individual state in \ref{fig:contr}. The ground state of $^{205}$Tl has the 
spin-parity $1/2^+$ so that Gamow-Teller type of transitions are possible only 
to $1/2^+$ and $3/2^+$ states in $^{205}$Pb. There are only three known $1/2^+$ 
states in $^{205}$Pb at 2795 keV, 4016 keV, and 4055 keV and no known $3/2^+$ states. 
GT transitions are therefore available only for $^8$B and $hep$ neutrinos, 
which happen to have relatively low fluxes in the standard solar 
models \cite{Bahcall2005}. For these higher-energy neutrinos the $1/2^+$ state 
at 4016 keV gives a noticeable contribution. Due to this lack of positive-parity 
states with small angular momenta, the cross section is dominated by spin-dipole 
type of forbidden transitions which render the total cross section smaller than
expected from energy arguments alone.

\section{Reaction rates in SNU}\label{sec:rates}

Owing to the small cross section of low-energy neutrino-nucleus interactions, it 
is convenient to present the neutrino-capture rate in the solar-neutrino units (SNUs), 
given as 1 SNU = $10^{-36}$ capture reactions per target atom per second.
Then the neutrino-capture rate $R$ is described by
 \begin{equation}
 \label{snu}
 R = 10^{36} \sum_i \int \sigma (E) \phi_i (E) dE \,,
 \end{equation}
where $E, \sigma  (E)$ and $\phi_i (E)$ are the neutrino energy, the neutrino-capture 
cross section (see the previous section) and the differential neutrino spectra. The 
latter are given at a distance of 1 astronomical unit (AU). The sum in (\ref{snu})
includes all 8 neutrino components from the pp and CNO cycle.

For the calculation of solar-neutrino capture rates the fluxes of the solar model 
BS05(OP) were adopted (see Table 2 in \cite{Bahcall2005}) with the neutrino spectrum 
shapes available on John N. Bahcall's website \cite{Bweb}. With these spectra and 
fluxes we get for the capture rate 100.2 SNU with $g_{\rm A}=0.75$ and 132.4 SNU 
with $g_{\rm A}=1.00$ when the survival probability of electron neutrinos is not taken 
into account. With the electron-neutrino survival probability of 0.54 for the pp, $^{7}$Be,
and $^{13}$N neutrinos, and 0.50 for the pep, $^{15}$O, and $^{17}$F neutrinos, and 0.36 
for the $^{8}$B neutrinos \cite{Borexino} we end up with a result of 
\begin{equation}
R(^{205}\rm Tl) = 62.2\pm8.6 \ \text{SNU} 
\end{equation}
for the capture rate.

\section{Summary and conclusions}

Given the revived interest in solar-neutrino detection using \iso{Tl}{205} due to its 
very small energy threshold we have performed a large-scale shell-model calculation 
to find out the cross section for the conversion of \iso{Tl}{205} to \iso{Pb}{205}. 
Combined with the neutrino fluxes predicted by established solar models and taking into
account the survival probabilities of electron neutrinos 
the capture rate turns out to be $ 62.2\pm8.6$ SNU. This capture rate is significantly smaller 
than anticipated by previous estimations.

\section{Acknowledgement} 

The authors would like to thank Dr. Y. Litvinov to bring this issue to our attention.  
This work has been partially supported by the Academy of Finland under the Academy 
project no. 318043. J. K. acknowledges the financial support from Jenny and Antti
Wihuri Foundation.

\end{document}